\documentclass[]{emulateapj}
\usepackage{natbib}
\usepackage{epsfig}

\newcommand{\rhoion}{\dot{n}_{\mathrm{ion}}} 
\newcommand{\fesc}{f_{\mathrm{esc}}}            
\newcommand{\xiion}{\xi_{\mathrm{ion}}}         
\newcommand{\rhoSFR}{\rho_{\mathrm{SFR}}}       
\newcommand{\s}{\mathrm{s}}                     
\newcommand{\yr}{\mathrm{yr}}                   
\newcommand{\Mpc}{\mathrm{Mpc}}                   
\newcommand{\Msun}{M_{\sun}}                    
\newcommand{\Lsun}{L_{\sun}}                    
\newcommand{\Lmin}{L_{\mathrm{min}}}            
\newcommand{\Lstar}{L_{\star}}                  
\newcommand{\ave}[1]{{\langle #1\rangle}}
\newcommand{\zreion}{z_{\mathrm{reion}}}         

\shorttitle{New Constraints on Cosmic Reionization}
\shortauthors{Robertson et al.}
\submitted{}

\begin{document}
\title{Cosmic Reionization and Early Star-Forming Galaxies: A Joint Analysis of New Constraints from Planck and Hubble Space Telescope}
\author{Brant E. Robertson\altaffilmark{1}, Richard S. Ellis\altaffilmark{2}, Steven R. Furlanetto\altaffilmark{3} and James S. Dunlop\altaffilmark{4}}
\altaffiltext{1}{Department of Astronomy, University of Arizona, Tucson, AZ 85721}
\altaffiltext{2}{Cahill Center for Astronomy and Astrophysics, California Institute of Technology, MS 249-17, Pasadena, CA 91125}
\altaffiltext{3}{Department of Physics \& Astronomy, University of California, Los Angeles CA 90095}
\altaffiltext{4}{Institute for Astronomy, University of Edinburgh, Edinburgh EH9 3HJ, UK}
\email{Email: brant@email.arizona.edu}

\begin{abstract}
We discuss new constraints on the epoch of cosmic reionization and test the assumption that most
of the ionizing photons responsible arose from high redshift star-forming galaxies. Good progress
has been made in charting the end of reionization through spectroscopic studies of $z\simeq$6-8
QSOs, gamma-ray bursts and galaxies expected to host Lyman $\alpha$ emission. However, the
most stringent constraints on its duration have come from the integrated optical depth, $\tau$, of Thomson scattering 
to the cosmic microwave background. Using the latest data on the abundance and 
luminosity distribution of distant galaxies from Hubble Space Telescope imaging, we simultaneously 
match the reduced value $\tau=0.066 \pm 0.012$ recently reported by the Planck 
collaboration and the evolving neutrality of the intergalactic medium with a reionization history 
within $6\lesssim z \lesssim 10$,  thereby reducing the requirement for a significant population of very
high redshift ($z\gg10$) galaxies. Our analysis strengthens the conclusion that star-forming galaxies 
dominated the reionization process and has important implications for upcoming 21cm experiments 
and searches for early galaxies with James Webb Space Telescope. 
\end{abstract}


\section{Introduction}
\label{sec:intro}

Cosmic reionization represents an important era for assembling a coherent picture of the
evolution of the Universe, and ambitious observational facilities are being constructed to 
explore the most important redshift range $7<z<20$. Through the Gunn-Peterson effect in 
high redshift QSOs and gamma ray bursts
\citep[GRBs, e.g.,][]{fan2006a,bolton2011a,chornock2013a,mcgreer2015a} and
the declining visibility of Lyman alpha (Ly$\alpha$) emission in high redshift galaxies 
\citep{stark2010a,pentericci2011a,pentericci2014a,schenker2012a,schenker2014a,treu2013a,tilvi2014a},
observations indicate that reionization ended by redshift $z\simeq6$.
However, the onset and duration of the reionization process remain less certain. The most convincing 
constraint is provided by the integrated optical depth, $\tau$, of Thomson scattering to the cosmic microwave 
background (CMB). The Wilkinson Microwave Anisotropy Probe (WMAP) delivered a value $\tau=0.088\pm0.014$
which, in the simplest model, corresponds to `instantaneous'  reionization at $\zreion\simeq10.5\pm1.1$\citep{hinshaw2013a}. 
As a result, the WMAP result has been widely interpreted as implying that reionization began at $z\simeq15$ or 
even earlier \citep{bromm2011a,dunlop2013a}.

Important information on the duration of reionization can now be determined from the star formation
rate (SFR) history \citep[][hereafter MD14]{madau2014a}, since early star-forming galaxies most likely 
supply the ionizing photons \citep{robertson2010a,robertson2013a}. 
This conclusion followed the first measures of their abundance over $8<z<10$
from Hubble Space Telescope (HST) Ultra Deep Field (UDF) observations \citep{beckwith2006a,koekemoer2013a,illingworth2013a}.
With plausible assumptions, $6<z<8$ star-forming galaxies can keep the Universe substantially ionized \citep{robertson2013a,finkelstein2014a}. 

However, to match the WMAP value of $\tau$, \citet{robertson2013a}
also required a significant population of star-forming galaxies beyond a redshift  $z\simeq10$.
As a direct census of $z>10$ galaxies is not currently possible, 
studies have since focused on the rate of decline in abundance over $8<z<10$ with mixed conclusions 
\citep[c.f.,][]{oesch2012a,oesch2013a,ishigaki2015a,mcleod2014a}.  The requirement for a
significant contribution of ionizing photons from $z>10$ galaxies remains an important uncertainty
whose resolution is perceived as a major goal for the James Webb Space Telescope (JWST).

The \citet{planck2015a} has recently 
reported a significantly lower value of the optical depth, $\tau=0.066\pm0.012$, 
consistent with a reduced redshift of instantaneous reionization, $\zreion=8.8^{+1.2}_{-1.1}$. 
Here we determine the extent to which the Planck result reduces the need for significant star formation in the 
uncharted epoch at $z > 10$. To demonstrate this, we calculate the contribution of $6<z<10$ star-forming galaxies 
to the integrated value of $\tau$, using the latest HST data. We then examine the residual contribution of ionizing 
photons required from sources beyond $z\simeq$10 to match the new value of $\tau$ from Planck, phrasing 
these constraints in terms of the likely abundance of $z>10$ galaxies that JWST would see in a typical deep exposure. 

Throughout we use the AB magnitude system \citep{oke1974a}, errors represent $1-\sigma$
uncertainties, and all cosmological
calculations assume flatness and the most recent Planck cosmological parameters 
\citep[$h=0.6774$, $\Omega_m=0.309$, $\Omega_b h^{2}=0.02230$, $Y_p=0.2453$;][]{planck2015a}.

\pagebreak

\section{Contribution of $z<10$ Galaxies to Late Reionization}
\label{sec:z610}

\subsection{Cosmic Star Formation History}

If Lyman continuum photons from star-forming galaxies dominate the reionization process,
an accounting of the evolving SFR
density will provide a measure of the time-dependent cosmic ionization rate
\begin{equation}
\label{eqn:rho_ion}
\rhoion = \fesc \xiion \rhoSFR,
\end{equation}
\noindent
where $\fesc$ is the fraction of photons produced by stellar populations
that escape to ionize the IGM, $\xiion$ is the number of Lyman continuum photons per 
second produced per unit SFR for a typical stellar population, and 
$\rhoSFR$ is the cosmic SFR density. Following \citet{robertson2013a}, 
we adopt a fiducial escape fraction of $\fesc=0.2$ and, motivated by the rest-frame UV spectral energy 
distributions of $z\sim7-8$ galaxies \citep{dunlop2013b}, a fiducial Lyman continuum photon 
production efficiency of $\log_{10} \xiion = 53.14~[\mathrm{Lyc}~\mathrm{photons}~\s^{-1} \Msun^{-1}~\yr]$.
Somewhat larger values of $\xiion$ may also be acceptable \citep[e.g.,][]{topping2015a}.

The observed infrared and rest-frame UV luminosity functions (LFs) provide
a means to estimate $\rhoSFR$. We use the recent compilation of IR and UV LFs provided in Table 1 of
MD14 and references therein to compute luminosity densities $\rho_L$ to a minimum
luminosity of $\Lmin=0.001\Lstar$, where $\Lstar(z)$ is the
characteristic luminosity of each relevant LF parameterization (e.g., Schechter
or broken power law models)\footnote{We adopt this limit since it corresponds to $M_{\mathrm{max}}\approx-13$
at $z\sim7$, which \citet{robertson2013a} found was required to reionize the Universe by $z\sim6$.
It corresponds to $M_{\mathrm{max}} = M_{\star} + 7.5$.}. We supplement the MD14 compilation
by including $\rhoSFR$ values computed from the LF determinations at $z\sim8$ by \citet{schenker2013a}, 
at  $z\sim7-8$ by \citet{mclure2013a}, and estimates at $z\sim10$ by \citet{oesch2014a} and \citet{bouwens2014a}. 
We include new HST Frontier Fields LF constraints at $z\sim7$ by \citet{atek2014a} and at
$z\sim9$ by \citet{mcleod2014a}, incorporating cosmic variance estimates from \citet{robertson2014a}.
We also updated the MD14 estimates derived from the 
\citet{bouwens2012a} LFs at $z\sim3-8$ with newer measurements 
by \citet{bouwens2014a}. All data were converted to the adopted 
Planck cosmology.

\begin{figure}
\figurenum{1}
\includegraphics[width=3.5in]{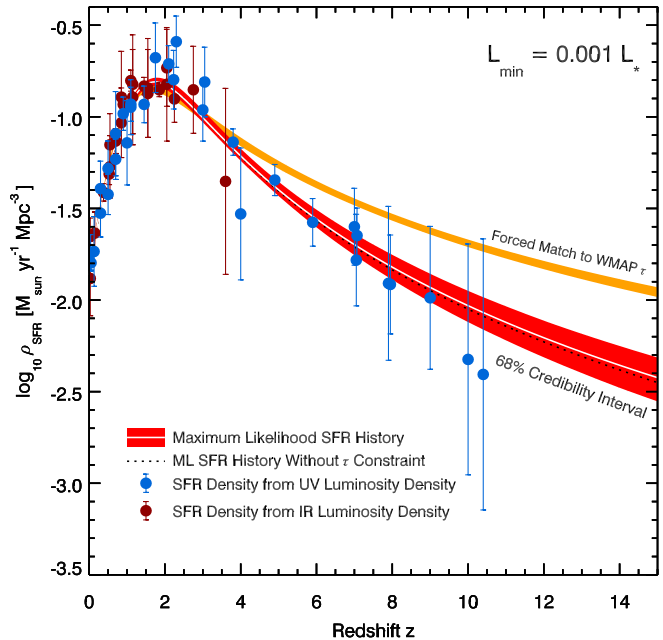}
\caption{\label{fig:sfrh}
Star formation rate density $\rhoSFR$ with redshift. Shown are the SFR
densities from \citet{madau2014a} determined from infrared (dark red points)
and ultraviolet (blue points) luminosity densities, updated for recent results
and extrapolated to a minimum luminosity $\Lmin=0.001\Lstar$.
A parameterized model for the evolving SFR density (Equation \ref{eqn:sfrh}) is
fit to the data under the constraint that the Thomson optical depth $\tau$ to electron
scattering measured by Planck is reproduced. The maximum likelihood model (white line)
and $68\%$ credibility interval on $\rhoSFR$ (red region)
are shown. A consistent SFR density history
is found even if the Planck $\tau$ constraint is ignored (dotted black line).
These inferences can be compared with a model forced to reproduce the previous
WMAP $\tau$ (orange region), which requires a much larger $\rhoSFR$ at redshifts $z>5$.
}
\end{figure}

We adopted the conversion $\rhoSFR = \kappa \rho_{L}$ supplied by MD14 for IR and
UV luminosity densities, i.e. $\kappa_{\mathrm{IR}}=1.73\times10^{-10}~\Msun~\yr^{-1}~\Lsun^{-1}$ 
and $\kappa_{\mathrm{UV}}=2.5\times10^{-10}~\Msun~\yr~\Lsun^{-1}$ respectively, as
well as their redshift-dependent dust corrections and a Salpeter initial mass function.
Uncertainties on $\rhoSFR$ are computed using 
faint-end slope uncertainties where available, and otherwise we increased the uncertainties reported
by MD14 by the ratio of the luminosity densities integrated to $L=0.03\Lstar$ and $L=0.001\Lstar$. 
The data points in Figure \ref{fig:sfrh} show the updated SFR densities and uncertainties 
determined from the IR (dark red) and UV (blue) LFs, each extrapolated to $\Lmin=0.001\Lstar$.

Since we are interested in the reionization history both up to and beyond the limit of the
current observational data, we adopt the four-parameter 
fitting function from MD14 to model $\rhoSFR(z)$,
\begin{equation}
\label{eqn:sfrh}
\rhoSFR(z) = a_p\frac{(1+z)^{b_p}}{1 + [(1+z)/c_p]^{d_p}}
\end{equation}
and perform a maximum likelihood (ML) determination of the
parameter values using Bayesian methods \citep[i.e., {\it Multinest};][]{feroz2009a} assuming Gaussian 
errors. If we fit to the data and uncertainties reported by MD14, we recover similar
ML values for the parameters of Equation \ref{eqn:sfrh}. 
The range of credible SFR histories can then be computed from the 
marginalized likelihood of $\rhoSFR$ by integrating over the full model parameter likelihoods.

\subsection{Thomson Optical Depth}

If photons from star forming galaxies drive the reionization
process, measures of the Thomson optical depth inferred from the CMB place additional constraints on $\rhoSFR$. The Thomson 
optical depth is given by
\begin{equation}
\label{eqn:tau}
\tau(z) = c\ave{n_{\mathrm{H}}}\sigma_\mathrm{T} \int_{0}^{z} f_{\mathrm{e}} Q_{\mathrm{H}_{\mathrm{II}}}(z')H^{-1}(z')(1+z')^2 dz'
\end{equation}
\noindent
where $c$ is the speed of light. The comoving hydrogen density 
 $\ave{n_{\mathrm{H}}}=X_{p}\Omega_{b}\rho_{c}$ involves 
 the hydrogen mass fraction $X_{p}$, the baryon density $\Omega_{b}$, and the critical density $\rho_{c}$.
The Thomson scattering cross section is $\sigma_{T}$. The number of free electrons per hydrogen
nucleus is calculated following \citet{kuhlen2012a} assuming doubly ionized helium at $z\le4$.

The IGM ionized fraction $Q_{\mathrm{H}_{\mathrm{II}}}(z)$ is computed
by evolving the differential equation
\begin{equation}
\label{eqn:QHII}
\dot{Q}_{\mathrm{H}_{\mathrm{II}}} = \frac{\rhoion}{\ave{n_{\mathrm{H}}}} - \frac{Q_{\mathrm{H}_{\mathrm{II}}}}{t_{\mathrm{rec}}}
\end{equation}
\noindent
where the IGM recombination time 
\begin{equation}
t_{\mathrm{rec}} = [C_{\mathrm{H}_{\mathrm{II}}} \alpha_{\mathrm{B}}(T)(1+Y_p/4X_p)\ave{n_{\mathrm{H}}}(1+z)^3]^{-1}
\end{equation}
\noindent
is calculated by evaluating the case B recombination coefficient $\alpha_{\mathrm{B}}$ at an IGM
temperature $T = 20,000$K and a clumping fraction $C_{\mathrm{H}_{\mathrm{II}}}=3$ \citep[e.g.,][]{pawlik2009a,shull2012a}. 
We incorporate the Planck Thomson optical depth constraints ($\tau = 0.066\pm0.012$, treated as a Gaussian) 
by computing the reionization history for
every value of the $\rhoSFR$ model parameters, evaluating Equation \ref{eqn:tau}, and then calculating the likelihood of
the model parameters given the SFR history data and the marginalized Thomson optical depth.

\begin{figure}
\figurenum{2}
\includegraphics[width=3.5in]{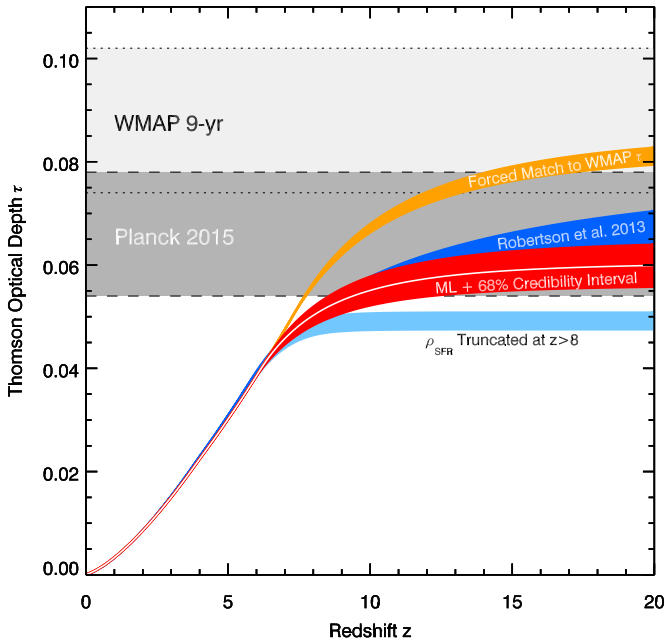}
\caption{\label{fig:tau}
Thomson optical depth to electron scattering $\tau$, 
integrated over redshift. Shown is the
Planck constraint $\tau=0.066\pm0.012$ (gray area), along with the 
marginalized $68\%$ credibility interval (red region)
computed from the SFR histories $\rhoSFR$
shown in Figure \ref{fig:sfrh}. 
The corresponding inferences of $\tau(z)$ from \citet{robertson2013a}
(dark blue region), a model forced to reproduce the 9-year WMAP 
$\tau$ constraints (orange region), and a model with $\rhoSFR$
truncated at $z>8$ (light blue region) following \citet{oesch2014a} 
are shown for comparison.
}
\end{figure}

Figure \ref{fig:sfrh} shows the ML and $68\%$ credibility interval (red region) on
$\rhoSFR(z)$ given the $\rhoSFR$ constraints and the newly-reported Planck Thomson optical depth. We find 
the parameters of Equation \ref{eqn:sfrh} to be
$a_p = 0.01376\pm0.001~\Msun~\yr~\Mpc^{-3}$, $b_p=3.26\pm0.21$, $c_p=2.59\pm0.14$, and $d_p=5.68\pm0.19$.
Without the Thomson optical depth constraint, the values change by less than $1\%$.
These inferences can be compared with a SFR history (Figure \ref{fig:sfrh}, orange region)
forced to match the previous WMAP measurement ($\tau=0.088\pm0.014$) by upweighting the contribution of
the derived $\tau$ value relative to the $\rhoSFR$ data. The model's ML parameters
($a_p=0.01306$, $b_p=3.66$, $c_p=2.28$, and $d_p=5.29$) lie well outside the range of
models that reproduce jointly $\rhoSFR(z)$ and the Planck $\tau$.
Fitting to only data at $z>3$ or only independent data points at $z>6$ changes our
credibility intervals by $\sim25\%$.

We can now address the important question of the redshift-dependent contribution of galaxies
to the Planck $\tau=0.066\pm0.012$ in Figure \ref{fig:tau}. The red region 
shows a history which is consistent with the SFR densities shown in Figure \ref{fig:sfrh} given 
our simple assumptions 
for the escape fraction $\fesc$, early stellar populations, and the clumpiness of the IGM.  
Importantly, the reduction in 
$\tau$ by Planck (compared to WMAP) largely eliminates the tension between
$\rhoSFR(z)$ and $\tau$ that was discussed
by many authors, including \citet{robertson2013a}. 
That a SFR history consistent with
the $\rhoSFR(z)$ data easily reproduces the Planck $\tau$ strengthens the conclusions of 
\citet{robertson2013a} that the bulk
of the ionizing photons emerged from galaxies. 
Figure~\ref{fig:tau} shows that the observed galaxy population at $z < 10$ can easily reach the 68\% credibility intervals of $\tau$ with plausible assumptions about $f_{\rm esc}$ and $L_{\rm min}$. 
As a consequence, the reduced $\tau$ eliminates the need for very high-redshift ($z\gg10$) star formation (see section 3 below). We note the dust correction used in computing
$\rhoSFR$ at $z\sim6$ permits an equivalently lower $\fesc$ without
significant change in the derived $\tau$.
We note that to reach $\tau\gtrsim0.08$ given the $\rhoSFR(z)$ constraints requires
$\fesc\gtrsim0.3$ or  $C_{\mathrm{H}_{\mathrm{II}}}\lesssim1$.

Figure \ref{fig:tau} also shows $\tau(z)$ computed with the $9-$year WMAP $\tau$ marginalized likelihood as a constraint on the high-redshift SFR density (blue region; \citealt{robertson2013a}), which favored a relatively low $\tau\sim0.07$.
If, instead, 
the SFR density rapidly declines as $\rhoSFR\propto(1+z)^{-10.9}$ beyond $z\sim8$
as suggested by, e.g., \citet{oesch2014a}, the Planck $\tau$ is not reached (light blue region).
Lastly, if we force the model to reproduce the best-fit WMAP $\tau$ (orange region), 
the increased ionization at high redshifts requires a dramatic increase in the $z>7.5$ 
SFR (see Figure~\ref{fig:sfrh}) and poses difficulties in matching other data on the IGM ionization state, as we discuss next.

\begin{figure}
\figurenum{3}
\includegraphics[width=3.5in]{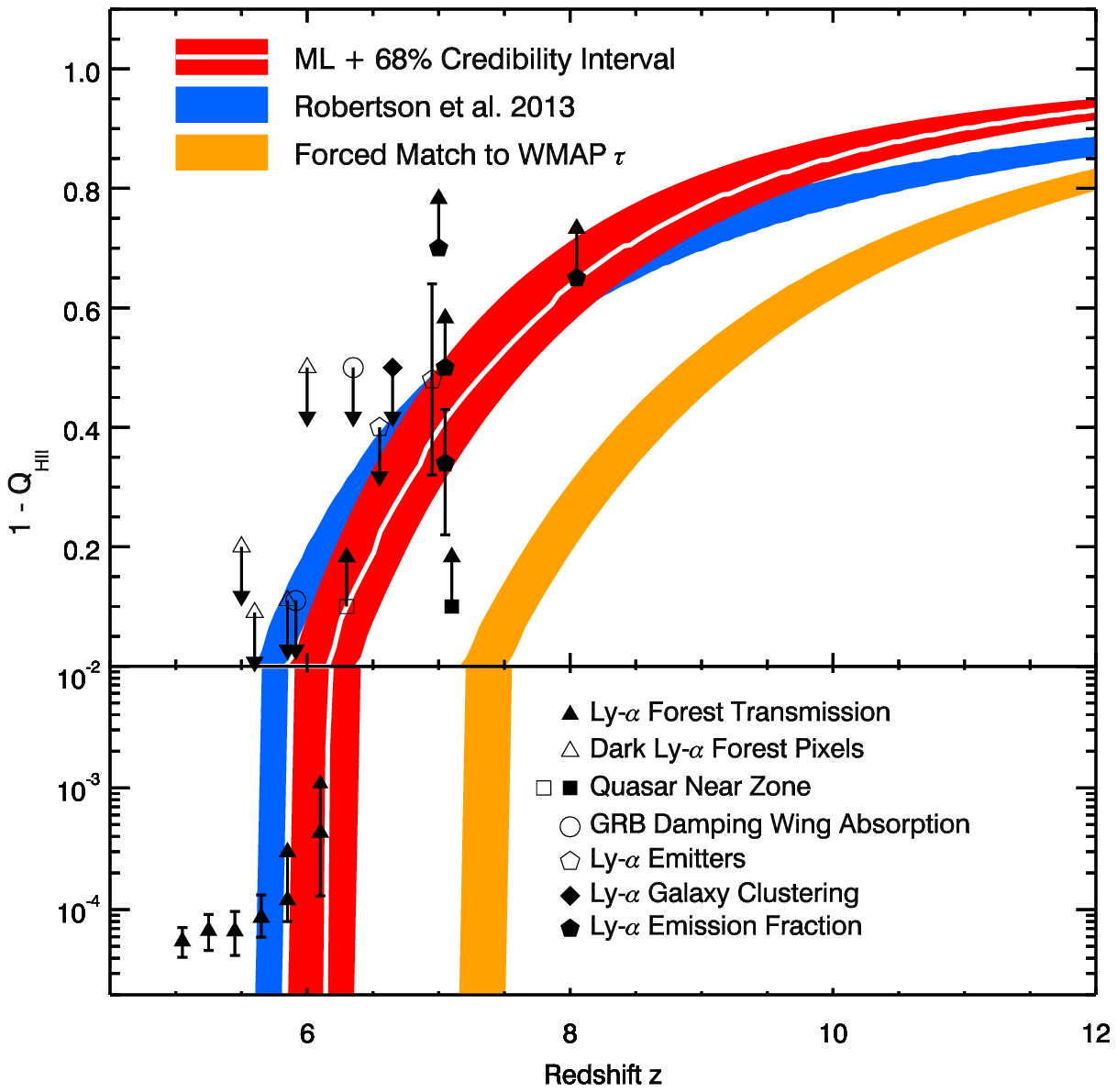}
\caption{\label{fig:QHI}
Measures of the neutrality $1-Q_{\mathrm{H}_\mathrm{II}}$ of the intergalactic
medium as a function of redshift. Shown are the observational constraints compiled
by \citet{robertson2013a}, updated to include recent IGM neutrality estimates
from the observed fraction of Lyman-$\alpha$ emitting galaxies \citep{schenker2014a,pentericci2014a},
constraints from the Lyman-$\alpha$ of GRB host galaxies \citep{chornock2013a}, and inferences
from dark pixels in Lyman-$\alpha$ forest measurements
\citep{mcgreer2015a}. The evolving IGM neutral fraction computed by the model is also shown (red region
is the $68\%$ credibility interval, white line is the ML model).
While these data are not used
to constrain the models, they are nonetheless remarkably consistent. The bottom panel shows the IGM
neutral fraction near the end of the reionization epoch, where the presented model fails to capture the
complexity of the reionization process.
For reference we also show the corresponding inferences calculated from \citet{robertson2013a} (blue region) and a model
forced to reproduce the WMAP $\tau$ (orange region). 
}
\end{figure}

\subsection{Ionization History}

Similarly, we can update our understanding of the evolving ionization fraction $Q_{\mathrm{H}_\mathrm{II}}(z)$ 
computed during the integration of Equation \ref{eqn:QHII}. Valuable observational
progress in this area made in recent years exploits the fraction of star forming galaxies showing Lyman-$\alpha$ emission
 \citep[e.g.,][]{stark2010a} now  extended to $z\sim7-8$ from \citet{treu2013a}, \citet{pentericci2014a} and
\citet{schenker2014a}, the Lyman-$\alpha$ damping wing absorption constraints from GRB
host galaxies by \citet{chornock2013a}, and the number of dark pixels in Lyman-$\alpha$
forest observations of background quasars \citep{mcgreer2015a}. While most of these results
require model-dependent inferences to relate observables to $Q_{\mathrm{H}_\mathrm{II}}$,
they collectively give strong support for reionization ending rapidly near $z\simeq$6.

Figure \ref{fig:QHI} shows these constraints, along with the
inferred $68\%$ credibility interval (red region; ML model shown in white)
on the marginalized distribution of the neutral fraction $1-Q_{\mathrm{H}_\mathrm{II}}$
from the SFR histories shown in Figure \ref{fig:sfrh} and the Planck
constraints on $\tau$. 
Although our model did not use these observations to constrain the computed reionization history,
we nonetheless find good agreement\footnote{The model does not fare well in comparison to Lyman-$\alpha$ 
forest measurements when 
$Q_{\mathrm{H}_\mathrm{II}}\sim1$ because of our simplified treatment of the ionization process \citep[see the discussion in][]{robertson2013a}}. 

Figure \ref{fig:QHI} also shows the earlier model of \citet{robertson2013a} (blue region) which completes reionization at slightly
lower redshift and displays a more prolonged ionization history. This model was in some tension with the
WMAP $\tau$ (Figure \ref{fig:tau}). If we force the model to reproduce
the WMAP $\tau$ (orange region), reionization ends by $z\sim7.5$, which is quite
inconsistent with several observations that indicate neutral gas within IGM  over the range $6\lesssim z \lesssim8$ 
(Figure \ref{fig:QHI}).

\begin{figure}
\figurenum{4}
\includegraphics[width=3.5in]{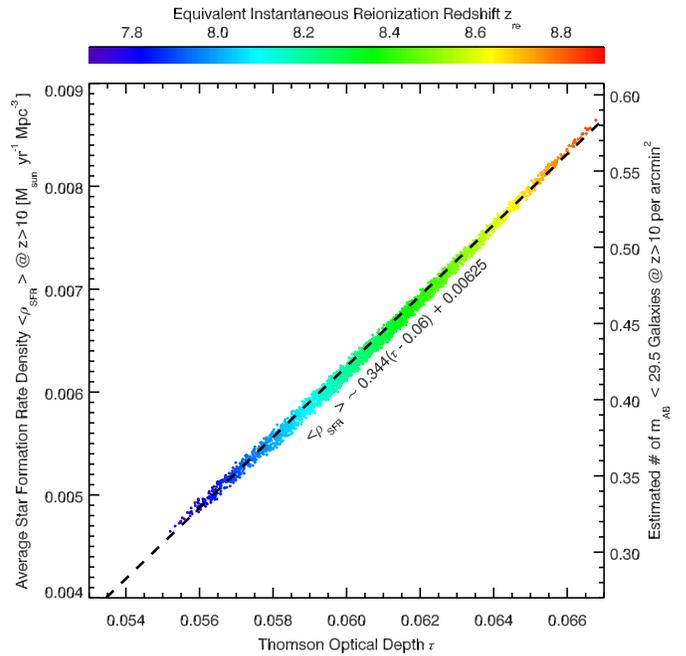}
\caption{\label{fig:tau_vs_sfr}
Correspondence between the Thomson optical depth, the equivalent instantaneous reionization
redshift 
$\zreion$, and the average SFR density $\rhoSFR$ at
redshift $z\gtrsim10$. Shown are samples (points) from the likelihood function of the $\rhoSFR$
model parameters resulting in the $68\%$ credibility interval on $\tau$ from Figure \ref{fig:tau},
color coded by the value of $\zreion$. The samples follow a tight, nearly linear correlation
(dashed line), demonstrating that in this model $\tau$ is a proxy for the high-redshift $\rhoSFR$.
We also indicate the number of $z>10$ galaxies
with $m_{\mathrm{AB}} < 29.5$ per arcmin$^{-2}$ (right axis), 
assuming the LF shape does not evolve above $z>10$.
}
\end{figure}

\section{Constraints on the Contribution of $z>10$ Galaxies to Early Reionization}

By using the parameterized model of MD14 to fit the cosmic SFR histories,
and applying a simple analytical model of the reionization process, we have
demonstrated that SFR histories consistent with the observed $\rhoSFR(z)$
integrated to $\Lmin=0.001\Lstar$ reproduce the observed Planck $\tau$ while
simultaneously matching measures of the IGM neutral fraction at redshifts
$6\lesssim z \lesssim 8$. As Figure \ref{fig:sfrh} makes apparent, the parameterized
model extends the inferred SFR history to $z>10$, beyond the reach of
current observations. Correspondingly, these galaxies supply a non-zero rate of
ionizing photons that enable the Thomson optical depth to slowly increase beyond
$z\sim10$ (see Figure \ref{fig:tau}). We can therefore ask whether a connection exists between 
$\rhoSFR(z>10)$ and the observed value of $\tau$ under the assumption that
star forming galaxies control the reionization process.

Figure \ref{fig:tau_vs_sfr} shows samples from the likelihood function of our model parameters
given the $\rhoSFR(z)$ and $\tau$ empirical constraints that indicate the mean SFR
density $\ave{\rhoSFR}$ (averaged over $10\lesssim z \lesssim 15$) as a function of the
total Thomson optical depth $\tau$. The properties $\ave{\rhoSFR}$ and $\tau$ are tightly
related, such that the linear fit
\begin{equation}
\ave{\rhoSFR} \approx 0.344(\tau - 0.06) + 0.00625~[\Msun~\yr^{-1}~\Mpc^{-3}]
\label{eq:z10fit}
\end{equation}
\noindent
provides a good description of their connection (dashed line). For reference,
the likelihood samples
shown in Figure \ref{fig:tau_vs_sfr} indicate the corresponding redshift of
instantaneous reionization $\zreion$ via a color coding.

Given that the SFR density is supplied by galaxies that
are luminous in their rest-frame UV, we can also connect the observed $\tau$
to the abundance of star forming galaxies at $z\gtrsim10$. This quantity holds
great interest for future studies with James Webb Space Telescope, as
the potential discovery and verification of distant galaxies beyond $z>10$
has provided a prime motivation for the observatory. The $5$-$\sigma$
sensitivity of JWST at $2~\mu$m in a $t=10^{4}$~s exposure is 
$m_{\mathrm{AB}}\approx29.5$.\footnote{See http://www.stsci.edu/jwst/instruments/nircam/sensitivity/table}
At $z\sim10$, this sensitivity corresponds to a UV absolute magnitude of $M_{\mathrm{UV}}\approx-18$.
Extrapolating the SFR density to $z>10$ and using the shape of the LF
at $z\ge9$, we estimate that $N\sim0.5~\mathrm{arcmin}^{-2}$ galaxies at $z>10$ will
be present at apparent magnitudes of $m_{\mathrm{AB}}<29.5$ at $\lambda=2~\mu$m. Deep observations
with JWST over $\sim10~\mathrm{arcmin}^{2}$ may therefore find $\gtrsim5$ candidates at $z>10$
\citep[see also][]{behroozi2015a}.
Returning to Figure \ref{fig:sfrh}, we can see the impact of the reduced value of $\tau$ by
comparing the Planck and WMAP curves beyond $z\simeq10$.

\section{Discussion}
\label{sec:discussion}

The lower value of the optical depth $\tau$ of Thomson scattering reported by the Planck consortium (2015) 
strengthens the likelihood that early star-forming galaxies dominated the reionization process, as our model can 
simultaneously match the observed SFR history (Figure \ref{fig:sfrh})
over $6<z<10$, the integrated value of $\tau$ (Figure \ref{fig:tau}), and recent constraints
on the IGM neutral fraction over $z\simeq6-8$ (Figure \ref{fig:QHI}). 

A state-of-the-art reionization analysis by \citet{choudhury2014a} used the distribution of 
Lyman $\alpha$ equivalent widths, the IGM photoionization rate, and the mean free path 
of ionizing photons, to also conclude that reionization likely completed at $z\sim6$, with a corresponding 
$\tau\approx0.07$\citep[see also][]{robertson2013a}. With Planck now favoring $\tau\approx0.066$ and 
informed by a full accounting of available constraints on the SFR history, we have reached similar
conclusions using different empirical inputs.

Our modeling makes some simplifying assumptions, adopting a constant escape fraction
$\fesc=0.2$, IGM clumping factor $C\approx3$, and Lyman continuum production efficiency
for early stellar populations. In \citet{robertson2013a} we examined these assumptions
carefully and tested more complex models, e.g. with evolving escape fraction required to match the
IGM photoionization rates at $z<6$ \citep[e.g.,][]{becker2013a}. These assumptions influence
the computation of $\tau$ and $Q_{\mathrm{H}_\mathrm{II}}$ but do not affect the inferred 
SFR history in Figure \ref{fig:sfrh}. Our conclusion that $z\lesssim10$ galaxies can account for the Planck $\tau$ 
relies on extrapolating LFs  below observed limits and a higher escape fraction than at lower redshift.  
If galaxies are less efficient ionizers, more $z>10$ star formation would be permitted. However, \citet{robertson2013a} 
already demonstrated such an ionizing efficiency is required to maintain a highly ionized IGM at $z \sim 7$ (Figure \ref{fig:QHI}). 

The ``excess" value of $\tau$ above that provided by galaxies at $z < 10$ measures
$\rho_{\rm SFR}$ at $z>10$.  Equation~\ref{eq:z10fit} 
and the Planck 1-$\sigma$ upper limit on $\tau$ provide 
an upper limit of $\rho_{\rm SFR} (z>10) \lesssim 0.013 \, M_\odot$~yr$^{-1}$~Mpc$^{-3}$.  This 
provides the first empirical limit on models that increase the ionizing efficiency during this
epoch e.g. with massive Population III stars and 
star formation in mini-halos (see \citealt{loeb2013a} for an overview of such models). Our results 
suggest such models cannot dramatically change the star formation efficiency at the earliest times.

Reionization proceeds relatively quickly as the ionized fraction evolves from
$Q_{\mathrm{H}_\mathrm{II}}=0.2$ to 
$Q_{\mathrm{H}_\mathrm{II}}=0.9$ in only $400~\mathrm{Myr}$
of cosmic history over $6\lesssim z \lesssim9$. This duration is
consistent with recent upper limits on the kinetic Sunyaev-Zel'dovich effect 
\citep[e.g.,][]{george2015a}. 
Our results offer extra hope for efforts to make redshifted 21-cm measurements of neutral hydrogen in the IGM, 
as the experimental foregrounds are weakest at low redshifts \citep[e.g.,][]{van_haarlem2013a,bowman2013a,pober2014a}. 
Such experiments are essential for testing key assumptions in our analysis (like $\fesc$ and $L_{\rm min}$) by observing 
the reionization process directly.  The apparent lateness of reionization suggests that next generation 
experiments, which hope to reach $z \sim 20$, can probe even earlier phases of galaxy formation.

\acknowledgments

We acknowledge useful discussions with George Efstathiou and Martin Haehnelt. 
This work was supported by Space Telescope Science Institute under award HST-GO-12498.01-A,
and the National Science
Foundation under Grant No. 1228509.
JSD acknowledges the support of the European
Research Council via the award of an Advanced Grant, and
the contribution of the EC FP7
SPACE project ASTRODEEP (Ref.No: 312725).

\end{document}